# Hierarchical Shape Abstraction for Analysis of Free-List Memory Allocators

Bin Fang[2,1] and Mihaela Sighireanu[1]

[1] IRIF, University Paris Diderot and CNRS, France
[2] Shanghai Key Laboratory of Trustworthy Computing, ECNU, China

**Abstract.** We propose a hierarchical abstract domain for the analysis of free-list memory allocators that tracks shape and numerical properties about both the heap and the free lists. Our domain is based on Separation Logic extended with predicates that capture the pointer arithmetics constraints for the heap-list and the shape of the free-list. These predicates are combined using a hierarchical composition operator to specify the overlapping of the heap-list by the free-list. In addition to expressiveness, this operator leads to a compositional and compact representation of abstract values and simplifies the implementation of the abstract domain. The shape constraints are combined with numerical constraints over integer arrays to track properties about the allocation policies (best-fit, first-fit, etc). Such properties are out of the scope of the existing analyzers. We implemented this domain and we show its effectiveness on several implementations of free-list allocators.

## 1 Introduction

A dynamic memory allocator (DMA) is a piece of software managing a reserved region of the heap. It appears in general purpose libraries (e.g., C standard library) or as part of applications where the dynamic allocation shall be controlled to avoid failure due to memory exhaustion (e.g., embedded critical software). A client program interacts with the DMA by requesting blocks of memory of variable size that it may free at any time. To offer this service, the DMA manages the reserved memory region by partitioning it into arbitrary sized blocks of memory, also called *chunks*. When a chunk is allocated to a client program, the DMA can not relocate it to compact the memory region (like in garbage collectors) and it is unaware about the kind (type or value) of data stored. The set of chunks not in use, also called *free chunks*, is managed using different techniques. In this paper, we focus on *free-list allocators* [19,26], that records free chunk in a list. This class of DMA includes textbook examples [17,19] and real-world allocators [20].

The automated analysis of DMA faces several challenges. Although the code of DMA is not long (between one hundred to a thousand LOC), it is highly optimised to provide good performance. Low-level code (e.g., pointer arithmetics, bit fields, calls to system routines like sbrk) is used to manage efficiently (i.e., with low additional cost in memory and time) the operations on the chunks in the reserved memory region. At the same time, the free-list is manipulated using high level operations over typed memory blocks (values of C structures) by mutating pointer fields without pointer arithmetic. The analyser has to deal efficiently with this *polar usage of the heap* made by the DMA. The invariants maintained by the DMA are complex. The memory region is organised into a *heap-list* based on the size information stored in the chunk header such that chunk overlapping and memory leaks are avoided. The start addresses of chunks shall be aligned to the some given constant. The free-list may have complex shapes (cyclic, acyclic, doubly-linked) and may be sorted by the start address of chunks to ease free chunks coalescing at deallocation. A precise analysis shall keep track of both numerical and shape properties to infer specifications implying such invariants for the allocation and deallocation methods of the DMA.

These challenges have been addressed partially by several works in the last ten years [5, 23, 25]. Efficient numerical analyses has been designed to track address alignment and bit-fields [23]. The most important progress has been done by the analysis proposed by Calcagno et al [5]. It is able to track the free-list shape and the numerical properties of chunk start addresses due to an abstract domain built on an extension of Separation Logic (SL) [24] with numerical constraints and predicates specifying memory blocks. However, some properties of the heap and free-list can not be tracked, e.g., the absence of memory leaks or ordering of start addresses of free-chunks. Although the analysis in [25] does not concern DMA, it is the first to propose a hierarchical abstraction of the memory to track properties of linked data structures stored in static memory regions. However, this analysis can not track properties like address sorting of the high level data structures (here the free-list) stored in the memory region. Furthermore, its link with a logic theory is not clear. Thus, a precise, logic based analysis for the inference of properties of free-list DMA is still a challenge.

In this paper, we propose a static analysis that is able to infer the above complex invariants of DMA on both heap-list and free-list. We define an abstract domain which uses logic formulas to abstract DMA configurations. The logic proposed extends the fragment of symbolic heaps of SL with a hierarchical composition operator $\Supset$ to specify that the free-list cover partially the heap-list. This operator provides a hierarchical abstraction of the memory region under the DMA control: the low-level memory manipulations are specified at the level of the heap-list and propagated in a way controlled by the abstraction at the level of the free-list. The shape specification is combined with a fragment of an array theory to capture properties of chunks in lists, similar to [3]. This combination is done in an accurate way wrt logic by including sequences of chunk addresses in the inductive definitions of list segments. The main advantages and contributions of this work are (1) *high precision of the abstraction* which is able to capture complex properties of free-list DMA implementations, (2) *strong logical basis* allowing to infer invariants that may be used by other verification methods, and (3) *modularity* of the abstract domain permitting to reuse existing abstract domains for the analysis of linked lists with integer data.

## 2   Overview

Fig. 1 includes excerpts from our running example, a free-list DMA implementation proposed in [1]. The type HDR (Fig. 1 (a)) defines the informations stored by the DMA at the start of chunks. The size field stores the full size of the chunk (in blocks of sizeof(HDR) bytes) and is used by the heap-list to determine the start of the next chunk. The fnx field is valid only for free chunks (i.e., chunks in the free-list) and stores the start address of the next free chunk. To simplify the presentation, we added the ghost field isfree, to mark explicitly free chunks. The memory region managed by the DMA is between addresses stored by the global variables _hsta and _hend; they are initialised by minit using sbrk calls. The start of the free list is stored in frhd. An intuitive view of the concrete state of the DMA is shown in Fig. 1(d). The busy chunks are represented in grey. The "next chunk" relation in the heap-list defined by the size field is represented by the lower arrows; the upper arrows represent the "next free chunk" relation defined by the fnx field. Furthermore, other structural invariants should be maintained after each call of DMA methods: the heap-list shall be correct inside the memory region [_hsta,_hend), consecutive chunks of the heap-list are not both free (*early coalescing* policy), the free-list shall include only chunks of the heap-list, be acyclic and sorted by start address of chunks. The allocation method searches a chunk with size bigger than the requested nbytes; if the chunk is larger, it is split in two parts such that the last part (the end of the initial chunk) is allocated.

The goal of our analysis is to establish that, if the client uses correctly the DMA methods, these methods (i) preserve the above structural invariants and (ii) are memory safe. In particular, we analyse the DMA methods starting from a client program which initialises the DMA and then calls allocation and deallocation methods (see Sec. 5) in a correct way.

*Heap-list abstraction.* The concrete memory configurations managed by the DMA are abstracted by a first abstraction layer based on the *symbolic heap graphs* fragment of SL [9]. The fragment is parameterised by a set of predicates which capture the heap-list as follows:

- The predicate $\mathsf{blk}(X;Y)$, introduced in [5], specifies an untyped sequence of bytes between the symbolic addresses $X$ and $Y$. E.g., the configuration obtained at line 20 of minit is abstracted by $\mathsf{blk}(\mathtt{\_hsta};\mathtt{\_hend})$.
- The predicate $\mathsf{chd}(X;Y)$ specifies a memory block $\mathsf{blk}(X;Y)$ storing a value of type HDR; the fields of this value are represented by the symbolic variables $X.\mathtt{size}$, $X.\mathtt{fnx}$, and $X.\mathtt{isfree}$ respectively.
- The predicate $\mathsf{chk}(X;Y)$ specifies a chunk built from a chunk header $\mathsf{chd}(X;Z)$ followed by a block $\mathsf{blk}(Z;Y)$ such that the full memory occupied has size $Y - X = X.\mathtt{size} \times \mathtt{sizeof(HDR)}$.
- A well formed heap-list segment starting at address $X$ and ending before $Y$ is specified using the predicate $\mathsf{hls}(X;Y)[W]$. The inductive definition of this predicate (see Tab. 2). requires that chunks does not overlap or leave memory leaks. The variable $W$ registers the sequence of start addresses of

```
1  typedef struct hdr_s {                    28  void* malloc(size_t nbytes)
2    struct hdr_s *fnx;                      29  {
3    size_t size;                            30    HDR *nxt, *prv;
4    //@ghost bool isfree;                   31    size_t nunits =
5  }  HDR;                                   32      (nbytes+sizeof(HDR)-1)/sizeof(HDR) + 1;
6                                            33
7  static void *_hsta = NULL;                34    for (prv = NULL, nxt = frhd; nxt;
8  static void *_hend = NULL;                35         prv = nxt, nxt = nxt->fnx) {
9  static HDR *frhd = NULL;                  36      if (nxt->size >= nunits) {
10 static size_t memleft;                    37        if (nxt->size > nunits) {
11                                           38          nxt->size -= nunits;
12 void minit(size_t sz)                     39          nxt += nxt->size;
13 {                                         40          nxt->size = nunits;
14   size_t align_sz;                        41        } else {
15   align_sz = (sz+sizeof(HDR)-1)           42          if (prv == NULL)
16             & ~(sizeof(HDR)-1);           43            frhd = nxt->fnx;
17                                           44          else
18   _hsta = sbrk(align_sz);                 45            prv->fnx = nxt->fnx;
19   _hend = sbrk(0);                        46        }
20                                           47        memleft -= nunits;
21   frhd = _hsta;                           48        //@ghost nxt->isfree = false;
22   frhd->size = align_sz / sizeof(HDR);    49        return ((void*)(nxt + 1));
23   frhd->fnx = NULL;                       50      }
24   //@ghost frhd->isfree = true;           51    }
25                                           52    warning("Allocation Failed!");
26   memleft = frhd->size;                   53    return (NULL);
27 }                                         54  }
```

(a) Globals and initialisation        (b) Allocation

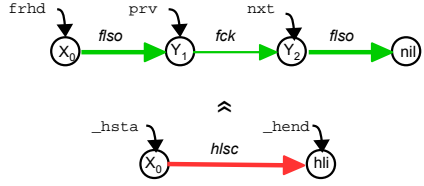   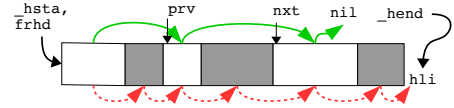

(c) Part of the abstract invariant at line 34        (d) Concrete memory

**Fig. 1.** Running example with code, concrete memory, and abstract specification

chunks in the list segment and it is used to put additional constraints on the fields of these chunks. For DMA with early coalescing of free-chunks, we abstract the heap-list segments by a stronger predicate, hlsc.

These predicates are combined using the *separation conjunction operator* ∗ of SL, which requires disjointness of memory regions specified by its operands. The bottom of Fig. 1(c) illustrates the heap-list abstraction of the concrete memory provided in Fig. 1(d); for readability, the abstraction is represented by its Gaifman graph. The ghost variable hli represents the end of the data segment of the DMA, as returned by `sbrk(0)`.

*Hierarchical abstraction of the free-list.* The first abstraction layer captures the total order of chunks in the heap-list. The free-list defines a total order over the set of free chunks. The second layer captures this order using the same SL fragment but over a different set of predicates (see Tab. 2):

- The predicate $\mathsf{fck}(X;Y)$ specifies a chunk $\mathsf{chk}(X;\ldots)$ starting at $X$, with $X.\mathtt{fnx}$ bound to $Y$ and $X.\mathtt{isfree}$ set to true.
- The predicate $\mathsf{fls}(X;Y)[W]$ specifies a free-list segment starting at $X$, whose last element field $\mathtt{fnx}$ points to $Y$; $W$ registers the sequence of start addresses of free chunks in the list segment. The predicate $\mathsf{flso}(X,\ldots)[W]$ abstracts free-list segments sorted by the start address of chunks.

The top of Fig. 1(c) illustrates the free-list abstraction by its Gaifman graph.

Finally, the memory shape abstraction is obtained by composing the two abstraction levels using a new operator $\Supset$, which requires that the set of chunks in the free-list abstraction is exactly the sub-set of chunks of the heap list with the $\mathtt{isfree}$ field set to true. Notice that the $\Supset$ operator can not be replaced by the logical conjunction because we are using the intuitive semantics of SL where spatial formulas fully specify the memory configurations. Or the free-list abstraction provides only a partial specification of the heap.

*Constraints over sequences of chunk addresses.* The inductive invariants specify invariants of DMA independent of user input, e.g., early chunk coalescing or free-list sorting. To capture policies like first-fit allocation (implemented by the $\mathtt{malloc}$ in Fig. 1(b)), we introduce universal constraints over sequences of chunk start addresses $W$ attached to shape atoms, like in [3]. For example, the first-fit policy obtained at line 37 of $\mathtt{malloc}$, is specified by:

$$\mathsf{hlsc}(X_0;\mathsf{hli})[W_H] \Supset (\mathsf{fls}(Y_0;Y_2)[W_1] * \mathsf{fck}(Y_2;Y_3) * \mathsf{fls}(Y_3;\mathsf{nil})[W_2]) \qquad (1)$$
$$\wedge\ Y_2.\mathtt{size} \geq \mathtt{nunits}\ \wedge\ \forall X \in W_1 \cdot X.\mathtt{size} < \mathtt{nunits}$$

where $Y_2$ is the symbolic address stored in the program variable $\mathtt{nxt}$. The general form of universal constraints is $\forall\ X \in W \cdot A_G \Rightarrow A_U$ where $A_G$ and $A_U$ are arithmetic constraints over $X$ and its fields. To obtain an efficient analysis, we fix $A_G$ and infer $A_U$. We require that both $A_G$ and $A_U$ belong to a class of constraints allowed by a numerical abstract domain (see Sec. 3).

*Static analysis with hierarchical shape abstraction.* Overall, the analysis algorithm is a standard shape analysis algorithm. To expose fields constrained or assigned by the program statements, it unfolds predicate definitions. To limit the size of the abstraction, the algorithm normalises formulas to maintain only symbolic addresses that are cut-points, i.e., they are stored in the program variables or are sharing points in lists. This transformation of formulas folds back sub-formulas into more general predicates. Because the set of normalised shape formulas is bounded, we define widening only for the sequence constraints.

The hierarchical shape requires to solve a number of specific issues (see Sec. 5). The unfolding of shape predicates shall be done at the appropriate level of abstraction. For example, a traversal of the free-list requires only unfolding and folding at the free-list level. The heap-list level may abstract chunks and symbolic addresses which are in the free-list level. Thus, we define protocols for the unfolding and folding operations at each level that are sound wrt the hierarchical composition $\Supset$ and with the sequence constraints.

**Table 1.** Logic syntax

| | |
|---|---|
| $X, Y \in \mathsf{AVar}$ location variables | $W \in \mathsf{SVar}$ sequence variables |
| $i, j \in \mathsf{IVar}$ integer variables | $\# \in \{=, \neq, \leq, \geq\}$ comparison operators |
| $x \in \mathsf{Var}$ logic variable | $\vec{x}, \vec{y} \in \mathsf{Var}^*$ vector of variables |
| $X.\mathtt{f}$ field access | $t, \Delta$ integer term resp. formula |
| $\varphi ::= \Pi \wedge \Sigma \mid \varphi \vee \varphi \mid \exists x \cdot \varphi$ | formulas |
| $\Pi ::= A \mid \forall X \in W \cdot A \Rightarrow A \mid W = w \mid \Pi \wedge \Pi$ | pure formulas |
| $A ::= X[.\mathtt{fnx}] - Y[.\mathtt{fnx}] \# t \mid \Delta \mid A \wedge A$ | location and integer constraints |
| $w ::= \epsilon \mid [X] \mid W \mid w.w$ | sequence terms |
| $\Sigma ::= \Sigma_H \ni \Sigma_F$ | spatial formulas |
| $\Sigma_H ::= \mathsf{emp} \mid \mathsf{blk}(X;Y) \mid \mathsf{chd}(X;Y) \mid \mathsf{chk}(X;Y) \mid X \mapsto x \mid$ $\mathsf{hls}(X;Y)[W] \mid \mathsf{hlsc}(X,i;Y,j)[W] \mid \Sigma_H * \Sigma_H$ | heap formulas |
| $\Sigma_F ::= \mathsf{emp} \mid \mathsf{fck}(X;Y) \mid \mathsf{fls}(X;Y)[W] \mid \mathsf{flso}(X,x;Y,y)[W] \mid \Sigma_F * \Sigma_F$ | free-list formulas |

## 3   Logic Fragment Underlying the Abstract Domain

We formalise in this section a fragment of Separation Logic [24] used to define the values of our abstract domain in Sec. 4.

*Syntax.* Let $\mathsf{AVar}$ be a set of *location variables* representing heap addresses; to simplify the presentation, we consider that $\mathsf{AVar}$ contains a special variable $\mathtt{nil}$ representing the null address, also denoted by $\mathtt{nil}$. Let $\mathsf{SVar}$ be a set of *sequence variables*, interpreted as sequences of heap addresses and $\mathsf{IVar}$ be a set of *integer variables*. The full set of *logic variables* is denoted by $\mathsf{Var} = \mathsf{AVar} \cup \mathsf{SVar} \cup \mathsf{IVar}$. The domain of heap addresses is denoted by $\mathbb{A}$ while the domain values stored in the heap is generically denoted by $\mathbb{V}$, thus $\mathbb{A} \subseteq \mathbb{V}$. To simplify the presentation, we fix $\mathtt{HDR}$, the type of chunk headers, and its fields $\{\mathtt{size}, \mathtt{fnx}, \mathtt{isfree}\}$ typed as declared in Fig. 1. The syntax of formulas is given in Tab. 1.

Formulas are in disjunctive normal form. Each disjunct is built from a pure formula $\Pi$ and a spatial formula $\Sigma$. Pure formulas $\Pi$ characterise the values of logic variables using comparisons between location variables, e.g., $X - Y = 0$, constraints $\Delta$ over integer terms, and sequence constraints. We let constraints in $\Delta$ unspecified, though we assume that they belong to decidable theories, e.g., linear arithmetic. The integer terms $t$ are built over integer variables and field accesses using classic arithmetic operations and constants. We denote by $\Pi_\forall$ (resp. $\Pi_W$, $\Pi_\exists$) the set of sub-formulas of $\Pi$ built from universal constraints (resp. sequence constraints, quantifier free arithmetic constraints).

A spatial formula has two components: $\Sigma_H$ specifies the heap-list and the locations outside this region; $\Sigma_F$ specifies only the free-list. The operator $\ni$ ensures that all locations specified by $\Sigma_F$ are start addresses of free chunks in the heap-list. The atom $\mathsf{emp}$ holds iff the domain of the heap is empty. The *points-to atom* $X \mapsto x$ specifies that the heap contains exactly memory block at location $X$ storing the value given by $x$. The *block atom* $\mathsf{blk}(X;Y)$ holds iff the heap contains a block of memory at location $X$ ending before the location $Y$. The other predicates are derived from $\mathsf{blk}$ and defined in Tab. 2. Notice that the *chunk header* atom $\mathsf{chd}(X;Y)$ does not expose the fields of the block at location $X$ using the points-to operator of SL. This ease the manipulation of heap-list level formulas, e.g., the coalescing of block and chunk atoms into a single block.

**Table 2.** Derived predicates

| |
|---|
| $\mathsf{chd}(X;Y) \triangleq \mathsf{blk}(X;Y) \wedge \mathtt{sizeof(HDR)} = Y - X \wedge X \equiv_{\mathtt{sizeof(HDR)}} 0$ |
| $\mathsf{chk}(X;Y) \triangleq \exists Z \cdot \mathsf{chd}(X;Z) * \mathsf{blk}(Z;Y) \wedge X.\mathtt{size} \times \mathtt{sizeof(HDR)} = Y - X$ |
| $\mathsf{fck}(X;Y) \triangleq \exists Z \cdot \mathsf{chk}(X;Z) \wedge X.\mathtt{isfree} = 1 \wedge X.\mathtt{fnx} = Y$ |
| $\mathsf{hls}(X;Y)[W] \triangleq \mathsf{emp} \wedge X = Y \wedge W = \epsilon$ |
| $\quad \vee\ \exists Z, W' \cdot \mathsf{chk}(X;Z) * \mathsf{hls}(Z;Y)[W'] \wedge W = [X].W'$ |
| $\mathsf{hlsc}(X, f_p; Y, f_\ell)[W] \triangleq \mathsf{emp} \wedge X = Y \wedge W = \epsilon \wedge 0 \leq f_p + f_\ell \leq 1$ |
| $\quad \vee\ \exists Z, W', f \cdot \mathsf{chk}(X;Z) * \mathsf{hlsc}(Z, f; Y, f_\ell)[W'] \wedge W = [X].W'$ |
| $\quad \wedge\ f = X.\mathtt{isfree} \wedge 0 \leq X.\mathtt{isfree} + f_p \leq 1$ |
| $\mathsf{fls}(X;Y)[W] \triangleq \mathsf{emp} \wedge X = Y \wedge W = \epsilon$ |
| $\quad \vee\ \exists Z, W' \cdot \mathsf{fck}(X;Z) * \mathsf{fls}(Z;Y)[W'] \wedge W = [X].W' \wedge X \neq Y$ |
| $\mathsf{flso}(X, x; Y, y)[W] \triangleq \mathsf{emp} \wedge X = Y \wedge W = \epsilon \wedge x - y \leq 0$ |
| $\quad \vee\ \exists Z, W' \cdot \mathsf{fck}(X;Z) * \mathsf{flso}(Z, X; Y, y)[W'] \wedge W = [X].W' \wedge x - X \leq 0$ |

*Semantics.* Formulas $\varphi$ are interpreted over pairs $(I, h)$ where $I$ is an *interpretation* of logic variables and $h$ is a *heap* mapping a location to a non-empty sequence of values stored at this location. Formally, $I \in [(\mathsf{AVar} \cup \mathsf{IVar}) \rightharpoonup \mathbb{V}] \cup [\mathsf{SVar} \rightharpoonup \mathbb{V}^*]$ and $h \in [\mathbb{A} \rightharpoonup \mathbb{V}^+]$ such that $\mathsf{nil} \notin \mathrm{dom}(h)$. Let $h(\ell)[i]$ denote the $i$th element of $h(\ell)$. Without loss of generality, we consider that a value of type HDR is a sequence of values indexed by fields. Tab. 3 provides the most important semantic rules. We denote by $[\![\varphi]\!]$ the pairs $(I, h)$ such that $I, h \models \varphi$. The semantic entailment $\varphi \Rightarrow \psi$ is defined by $[\![\varphi]\!] \subseteq [\![\psi]\!]$.

*Transformation rules.* The definitions in Tab. 2 imply a set of lemmas used to transform formulas in abstract values (in Sec. 5). The first set of lemmas is obtained by directing predicate definitions in both directions. For example, each definition $P(\ldots) \triangleq \vee_i \varphi_i$ introduces a set of *folding* lemmas $\varphi_i \Rightarrow P(\ldots)$ and an *unfolding* lemma $P(\ldots) \Rightarrow \vee_i \varphi_i$.

The second class of lemmas concerns list segment predicates in Tab. 2. The inductive definitions of these predicates satisfy the syntactic constraints defined in [12] for *compositional predicates*. Thus, every $P \in \{\mathsf{hls}, \mathsf{hlsc}, \mathsf{fls}, \mathsf{flso}\}$ satisfies the following *segment composition lemma*:

$$P(X, \vec{x}; Y, \vec{y})[W_1] * P(Y, \vec{y}; Z, \vec{z})[W_2] \wedge W = W_1.W_2 \quad \Rightarrow \quad P(X, \vec{x}; Z, \vec{z})[W] \quad (2)$$

**Table 3.** Logic semantics: main rules

| | | |
|---|---|---|
| $I, h \models \Sigma_H \ni \Sigma_F$ | iff | $I, h \models \Sigma_H$ and $\exists h' \subseteq h$ s.t. $I, h' \models \Sigma_F$ |
| | | $\forall \ell \in \text{dom}(h') \cdot h'(\ell)[\texttt{isfree}] = 1$ |
| $I, h \models \text{emp}$ | iff | $\text{dom}(h) = \varnothing$ |
| $I, h \models \text{blk}(X; Y)$ | iff | $\text{dom}(h) = I(X) \wedge I(Y) - I(X) = |h(I(X))|$ |
| $I, h \models X \mapsto x$ | iff | $\text{dom}(h) = I(X) \wedge h(I(X))[0] = I(x)$ |
| $I, h \models \Sigma_1 * \Sigma_2$ | iff | $\exists h_1, h_2$ s.t. $h = h_1 \uplus h_2$ and $I, h_i \models \Sigma_i$ for $i = 1, 2$ |
| $I, h \models \forall X \in W \cdot A_1 \Rightarrow A_2$ | iff | $I(W) = [a_1, \ldots, a_n]$ s.t. $\forall i \in (1..n)\ I[X \mapsto a_i], h \models A_1 \Rightarrow A_2$ |
| where | | |
| $h_1 \subseteq h_2$ | iff | $\text{dom}(h_1) \subseteq \text{dom}(h_2)$ and $\forall \ell \in \text{dom}(h_1) \cdot h_1(\ell) = h_2(\ell)$ |
| $h_1 \circledast h_2$ | iff | $\forall l_1 \in \text{dom}(h_1), l_2 \in \text{dom}(h_2) \cdot l_1 \neq l_2 \wedge$ |
| | | $\big((l_1..l_1 + |h_1(l_1)| - 1) \cap (l_2..l_2 + |h_2(l_2)| - 1)\big) = \varnothing$ |
| $h = h_1 \uplus h_2$ | iff | $h_1 \circledast h_2$, $\text{dom}(h) = \text{dom}(h_1) \uplus \text{dom}(h_2)$, and |
| | | $(h_1 \uplus h_2)(\ell) \triangleq \begin{cases} h_1(\ell) & \text{if } \ell \in \text{dom}(h_1) \\ h_2(\ell) & \text{if } \ell \in \text{dom}(h_2) \end{cases}$ |

The reverse implication is applied to split non-empty list segments. Finally, the block sub-formulas are removed, split, or folded using the following lemmas:

$$\text{blk}(X; Y) \wedge X \geq Y \Rightarrow \text{emp} \qquad (3)$$

$$\text{blk}(X; Y) \wedge X < Y \Rightarrow \text{blk}(X; Z) * \text{blk}(Z; Y) \wedge X \leq Z \leq Y \qquad (4)$$

$$\text{blk}(X; Y) * \text{blk}(Y'; Z) \wedge X \leq Y = Y' \leq Z \Rightarrow \text{blk}(X; Z). \qquad (5)$$

## 4 Abstract Domain for Hierarchical Shape Abstraction

We define in this section the join-semilattice $\langle \mathcal{A}, \sqsubseteq, \sqcup \rangle$ used in our analysis. It is parameterised by a numerical abstract domain $\langle \mathcal{N}, \sqsubseteq^\mathcal{N}, \sqcup^\mathcal{N} \rangle$.

*Concrete states.* Let $\mathbb{X}$ be the set of program variables, where hli is a ghost variable of location type. Values in $\mathcal{A}$ represent sets of concrete states $M \in \mathbb{M}$ of the program. A concrete state $M$ encloses an environment $\epsilon \in \mathbb{E} = \mathbb{X} \to \mathbb{A}$ mapping each program variable to its storing location, and a heap $h : \mathbb{A} \to \mathbb{V}^+$ mapping locations to sequences of values. For simplicity, hli symbol is overloaded to denote the symbolic location stored by hli.

*Abstract values.* Values in $\mathcal{A}$ are a restricted form of logic formulas. Generally speaking, $\mathcal{A}$ is a *co-fibered product* [6] of an *extended symbolic heap domain* for the spatial part and a *data word* domain [3] for the pure part. More precisely, $\mathcal{A}$ includes a special value for $\top$ and finite mappings of the form:

$$a^\sharp ::= \{\langle \epsilon_i^\sharp, \Sigma_i(\overrightarrow{x}, \overrightarrow{W})\rangle \mapsto \Pi_i(\overrightarrow{x}, \overrightarrow{W} \cup \{W_H, W_F\})\}_{i \in I} \qquad (6)$$

where $\epsilon_i^\sharp : \mathbb{X} \to \text{Var}$ is an abstract environment mapping program variables to symbolic location variables, $\Pi_i$ includes arithmetic constraints allowed by $\mathcal{N}$, and the free variables of each formulas are detailed. Furthermore, the usage of sequence variables in $\Sigma_i$ and $\Pi_i$ are restricted as follows:

**R$_1$:** A sequence variable is bound to exactly one list segment atom in $\Sigma_i$; thus $\Sigma_i$ defines an injection between list segment atoms and sequence variables.
**R$_2$:** $\Pi_i$ contains only the sequence constraints $W_H = w$ and $W_F = w'$, where $W_H$ and $W_F$ are special variables representing the full sequence of start addresses of chunks in the heap resp. free-list levels.

In addition, the universal constraints in the pure formulas $\Pi_i$ are restricted such that, in any formula $\forall X \in W \cdot A_G \Rightarrow A_U$:

**R$_3$:** $A_G$ and $A_U$ use only terms where $X$ appears inside a field access $X.\mathtt{f}$.
**R$_4$:** $A_G$ has one of the forms $X.\mathtt{size}\#i$ or $X.\mathtt{isfree} = i$.

These restrictions still permit to specify DMA policies like first-fit (see eq. (1)) and besides enable an efficient inference of universal constraints.

*Internal representation.* To ease the manipulation of extended spatial formulas $\langle \epsilon^\sharp, \Sigma \rangle$, we use the Gaifman graph representation, like in Fig. 1(c); nodes represent symbolic locations variables and labeled edges represent the spatial atoms in $\Sigma$ or mappings in $\epsilon^\sharp$. The universal formulas are represented by a map binding each pair of sequence variable and some guard $A_G$ to a numerical abstract value.

*Concretisation.* An abstract value of the form (6) represents a formula $\vee_i \exists \vec{x}, \vec{W} \cdot \Sigma_i \wedge \Pi_i \wedge \epsilon_i^\sharp$ where each binding $(v, x) \in \epsilon_i^\sharp$ is encoded by $v \mapsto x$ ($v$ is the location where is stored the program variable $v$). The false formula is represented by the empty mapping which corresponds to $\bot$. Based on the correspondence between abstract values and logic formulas, we define the concretisation $\gamma : \mathcal{A} \to \mathbb{M}$ by $\gamma(a^\sharp) = [\![a^\sharp]\!]$.

*Ordering.* The partial order $\sqsubseteq$ is defined using a sound procedure inspired by [4, 12]. For any two non trivial abstract values $a^\sharp, b^\sharp \in \mathcal{A}$, $a^\sharp \sqsubseteq b^\sharp$ if for each binding $\langle \epsilon_i^\sharp, \Sigma_i \rangle \mapsto \Pi_i \in a^\sharp$ there exists a binding $\langle \epsilon_j^\sharp, \Sigma_j \rangle \mapsto \Pi_j \in b^\sharp$ such that:

- there is a graph isomorphism between the Gaifman graphs of spatial formula at each level of abstraction from $\Sigma_i$ to $\Sigma_j$; this isomorphism is defined by a bijection $\Psi : \mathrm{img}(\epsilon_i^\sharp) \to \mathrm{img}(\epsilon_j^\sharp)$ between symbolic location variables and a bijection $\Omega$ between sequence variables. Thus, $\Sigma_i[\Psi][\Omega] = \Sigma_j$,
- for each sequence constraint $W = w$ in $\Pi_{W,i}$, $\Omega(W) = \Omega(w)$ is a sequence constraint in $\Pi_{W,j}$,
- $\Psi(\Pi_{\exists,i}) \sqsubseteq^\mathcal{N} \Pi_{\exists,j}$,
- for each $W$ defined in $\Sigma_i$ and for each universal constraint $\forall X \in W \cdot A_G \Rightarrow A_U$ in $\Pi_{\forall,i}$, then $\Pi_{\forall,j}$ contains a universal constraint on $W' = \Omega(W)$ of the form $\forall X \in W' \cdot A_G \Rightarrow A'_U$ such that $\Psi(\Pi_{\exists,i} \wedge A_U) \sqsubseteq^\mathcal{N} A'_U$.

The following theorem is a consequence of restrictions on the syntax of formulas used in the abstract values.

**Theorem 1 ($\sqsubseteq$ soundness).** *If $a^\sharp \sqsubseteq b^\sharp$ then $\gamma(a^\sharp) \subseteq \gamma(b^\sharp)$.*

*Join.* Given two non-trivial abstract values, $a^\sharp$ and $b^\sharp$, their join is computed by joining the pure parts of bindings with isomorphic shape graphs [3]. Formally, for each two bindings $\langle \epsilon_i^\sharp, \Sigma_i \rangle \mapsto \Pi_i \in a^\sharp$ and $\langle \epsilon_j^\sharp, \Sigma_j \rangle \mapsto \Pi_j \in b^\sharp$ such that there is a graph isomorphism defined by $\Psi$ and $\Omega$ between $\langle \epsilon_i^\sharp, \Sigma_i \rangle$ and $\langle \epsilon_j^\sharp, \Sigma_j \rangle$, we define their join to be the mapping $\{\langle \epsilon_j^\sharp, \Sigma_j \rangle \mapsto \Pi\}$ where $\Pi$ is defined by:

- $\Pi$ includes the sequence constraints of $b^\sharp$, i.e., $\Pi_W \triangleq \Pi_{W,j}$,
- $\Pi_\exists \triangleq \Psi(\Pi_{\exists,i}) \sqcup^{\mathcal{N}} \Pi_{\exists,j}$,
- for each $W$ sequence variable in $\mathrm{dom}(\Omega)$ and for each type of constraint $A_G$, then $\Pi_\forall$ contains the formula $\forall X \in \Omega(W) \cdot A_G \Rightarrow \Psi(A_{U,i}) \sqcup^{\mathcal{N}} A_{U,j}$ where $A_{U,i}$ (resp. $A_{U,j}$) is the constraint bound to $W$ (resp. $\Omega(W)$) in $\Pi_{\forall,i}$ (resp. $\Pi_{\forall,j}$) for guard $A_G$ or $\top$ if no such constraint exists.

The join of two bindings with non-isomorphic spatial parts is the union of the two bindings. Then, $(a^\sharp \sqcup b^\sharp)$ computes the join of bindings in $a^\sharp$ with each binding in $b^\sharp$. Intuitively, the operator collects the disjuncts of $a^\sharp$ and $b^\sharp$ but replaces the disjuncts with isomorphic spatial parts by one disjunct which maps the spatial part to the join of the pure parts. Two universal constraints are joined when they concern the same sequence variables and guard $A_G$ since $((\forall c \in W \cdot A_G \Rightarrow A_1) \vee (\forall c \in W \cdot A_G \Rightarrow A_2)) \Rightarrow (\forall c \in W \cdot A_G \Rightarrow (A_1 \vee A_2))$.

**Theorem 2 ($\sqcup$ soundness).** *For any $a^\sharp, b^\sharp \in \mathcal{A}$, $\gamma(a^\sharp) \cup \gamma(b^\sharp) \subseteq \gamma(a^\sharp \sqcup b^\sharp)$.*

*Cardinality of the abstract domain.* The number of mappings in (6) increases during the symbolic execution by the introduction of new existential variables keeping track of the created chunks. Although the analysis stores only values with linear shape of lists (other shapes are signalled as error), the number of linear shapes is exponential in the number of nodes, in general. We avoid this memory explosion by eliminating existential variables using the transformation rules that replace sub-formulas by predicates, an operation classically called *predicate folding*. This operation uses lemmas (2)–(5), as discussed in Sec. 5. Thus, the domain of abstract values is bounded by an exponential on the number of pointer program variables local to DMA methods which is small in general, e.g., $\leq 3$ in our benchmark. However, the domain of pure formulas used in the image of abstract values is not bounded because of integer constants. This fact requires to define widening operators for the data word domain used for the pure constraints.

## 5    Analysis Algorithm

We now describe the specific issues of the static analysis algorithm based on the hierarchical abstract domain presented in Sec. 4.

### 5.1    Main principles

The analysis is defined as a forward, non-relational abstract interpretation [8] over a shape abstract domain, and follows the principles of [7, 9, 10].

```
1  int main(void) {
2    minit(1024);
3    void* p = malloc(20);
4    malloc(20);
5    mfree(p); p = NULL;
6    p = malloc(20);
7    malloc(20);
8    mfree(p); p = NULL;
9    return 0;
10 }
```

**Fig. 2.** A client program

The DMA methods are analysed starting from some client function, like the main function in Fig. 2. The client programs are chosen to reveal policies of the DMA concerning the free chunk coalescing and the choice of the free chunk to be allocated. The analysis computes, for each control location $\ell$, an abstract value $a_\ell^\sharp$ such that the $\gamma(a_\ell^\sharp)$ includes all the program configurations reachable by the concrete semantics of the program. For this, we define sound abstract transformers to compute abstract post-conditions and widening operators to speed-up the convergence of the fix-point computation. The original points on abstract transformers concern the transfer of information between layers in the hierarchical unfolding, splitting, and folding of predicates, as detailed in Sec. 5.2–5.3. Furthermore, these operations are defined in a modular way, by extending [6] to data word numerical domains. The widening operator uses the widening of data word domain defined in [10].

### 5.2 Hierarchical unfolding

Abstract transfer functions compute over-approximations of post-images of atomic conditions and assignments in the program. For the spatial part, the abstract value is transformed such that the program variables read or written by the program operation are constrained using predicates that may capture the effect of the program operation. This transformation is called *predicate unfolding*.

We define the following partial order between predicates blk < chd < chk < fck < hls, hlsc, fls, flso which intuitively corresponds to an increasing degree of specialisation. For each program operation $s$ and each pointer variable x in $s$, an atom $P(X;\ldots)$ with $\epsilon^\sharp(\mathtt{x}) = X$ is transformed using lemmas in Sec. 3 to obtain the atom $Q(X;\ldots)$ such that $Q \leq P$ is the maximal predicate satisfying:

- if $s$ reads the fields of HDR, then $Q \leq$ fck,
- if $s$ assigns x.isfree or x.fnx, then $Q \leq$ chk,
- if $s$ mutates x using pointer arithmetics or assigns x.size, then $Q \leq$ chd.

Consider the condition `nxt->size > nunits` at line 37 in Fig. 1(b), which reads the field size. Applied to the abstract value in Fig. 1(c), it requires to unfold the flso predicate from $Y_2$, to obtain the top part of Fig. 3(a). To compute the post-image of the next assignment `nxt->size -= nunits`, the symbolic location $Y_2$ shall be the root a chd predicate (third case above). Thus, $Y_2$ is instantiated in the heap-list by first splitting and then unfolding the hlsc predicate, then by unfolding chk to obtain the bottom part of Fig. 3(a). The unfolding of chk requires to remove the fck atom from $Y_2$ in the free-list level because its definition is not more satisfied at the heap-list level.

The assignment `nxt += nxt->size` does not require to transform the predicate rooted in $Y_2$ because it is already $\leq$ chd. Instead, the transformer adds a new symbolic location $Z_1$ in the heap-list level and constrain it by $Z_1 = Y_2 + Y_2.\mathtt{size} \times \mathtt{sizeof(HDR)}$. If $Z_1$ goes beyond the limits of the user part of the chunk starting at $Y_2$ (i.e., outside the interval $[X_1, X_2]$ in Fig. 3(a)), the analysis signals a chunk breaking. Otherwise, the blk atom from $X_1$ is split using lemma (4) to insert $Z_1$; the result is given by the top part of Fig. 3(b).

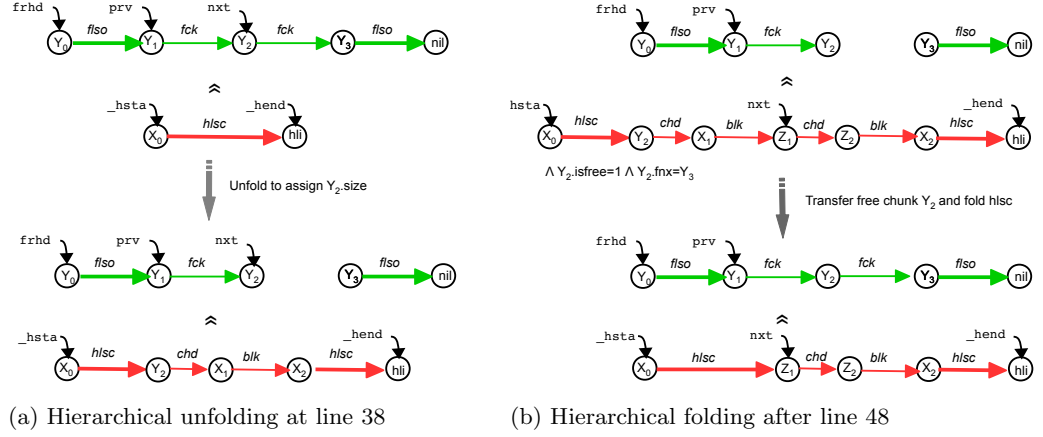

(a) Hierarchical unfolding at line 38    (b) Hierarchical folding after line 48

**Fig. 3.** Hierarchical abstract transformers for the running example

### 5.3  Hierarchical folding

To reduce the size of abstract values, the abstract transformers finish their computation on a binding $\langle \epsilon_i^\sharp, \Sigma_i \rangle \mapsto \Pi_i$ by eliminating the symbolic locations which are not cut-points in $\Sigma_i$. The elimination uses predicate folding lemma like (2) or (5) to replace sub-formulas using these variables by one predicate atom. The graph representation eases the computation of sub-formulas matching the left part of folding lemma.

More precisely, the elimination process has the following steps. First, it searches sequences of sub-formula of the form $\mathsf{chd}(X_0; X_1) * \mathsf{blk}(X_1; X_2) \ldots * \mathsf{blk}(X_{n-1}; X_n)$ where none of $X_i$ ($1 \le i < n$) is in $\mathsf{img}(\epsilon^\sharp)$. Such sub-formulas are folded into $\mathsf{chk}(X_0; X_n)$ if the pure part of the abstract value implies $X_0.\mathtt{size} \times \mathtt{sizeof(HDR)} = X_n - X_0$ (see Tab. 2). We use the variable elimination provided by the numerical domain $\mathcal{N}$ to project out $\{X_1, \ldots, X_{n-1}\}$ from the pure part. Furthermore, if the pure part implies $X_0.\mathtt{isfree} = 1$, then the chunk atom (and its start address) is promoted as fck to the free-list level. This step is illustrated in Fig. 3(b) by sub-formulas $\mathsf{chd}(Y_2; X_1) * \mathsf{blk}(X_1; Z_1)$.

The second step folds list segments by applying their inductive definition and the composition lemma (2). The atoms $\mathsf{chk}(X_0; \ldots)$ for which the free-list level contains an atom $\mathsf{fck}(X_0; \ldots)$ may be folded at the heap-level into list segments due to the semantics of $\Rrightarrow$. For example, the chunk starting from locations $Y_2$ is folded inside the heap-list segment in Fig. 3(b). Notice that folding of list segments implies the update of sequence and universal constraints like in [10].

Table 4. Benchmark of DMA

| DMA | | LOC | List Pred. | $|a^\sharp|$ | $|W_H|/|W_F|$ | Invariants |
|---|---|---|---|---|---|---|
| DK$_{\text{FF}}$ | [19] | 176 | hlsc, flso | 25 | 8/5 | first-fit, `MIN_SIZE`-size |
| DK$_{\text{BF}}$ | [19] | 130 | hlsc, flso | 26 | 8/6 | best-fit, `MIN_SIZE`-size |
| LA | [1] | 181 | hlsc, flso | 25 | 8/5 | first-fit, 0-size |
| DK$_{\text{NF}}$ | [19] | 137 | hlsc, flso | 30 | 8/6 | first-fit, `MIN_SIZE`-size |
| KR | [17] | 284 | hlsc, flso | 32 | 8/6 | first-fit, 0-size |

## 6 Experiments

We implemented the abstract domain in Ocaml and integrated it into an analyser as plug-in[3] of the Frama-C platform [18]. We are using several modules of Frama-C, e.g., C parsing, abstract syntax tree transformations, and fix-point computation. The data word domain uses as numerical domain $\mathcal{N}$ the library of polyhedra with congruence constraints provided by Apron [16]. To obtain precise numerical invariants, we transform program statements using bit-vector operations (e.g., line 16 of Fig. 1(a)) into statements permitted by the polyhedra domain that over-approximate the original effet.

We applied our analysis on the benchmark of free-list DMA in Tab. 4. DK$_{\text{FF}}$ and DK$_{\text{BF}}$ are implementations of Algorithms A and B from Sec. 2.5 of [19]. These DMA keep an acyclic free-list sorted by the start addresses of chunks. The deallocation does coalescing of successive free chunks. The allocation implements a first-fit resp. best-fit policy such that the fitting chunk is not split if the remaining free part is less than `MIN_SIZE` (variant proposed in [19]). This property is expressed by the invariant "`MIN_SIZE`-size" $\forall X \in W_H \cdot X.\texttt{size} \geq 8$ (here `MIN_SIZE` is 8 bytes) which is inferred by our analyser. The best-fit policy is implied by the universal constraint $\forall X \in W_i \cdot X.\texttt{size} \geq \texttt{rsz} \Rightarrow X.\texttt{size} \geq Y.\texttt{size}$ where rsz is the requested size, $Y$ is the symbolic address of the fitting chunk, and $W_i$ represents list segments around the fitting chunk. LAis our running example in Fig. 1; it follows the same principles as DK$_{\text{FF}}$, but get rid of the constraint for chunk splitting. Our analyser infers the "0-size" invariant, i.e., $\forall X \in W_H \cdot X.\texttt{size} \geq 4$ (=`sizeof(HDR)`). Notice that the code analysed fixes an obvious problem of the `malloc` method published in [1]. DK$_{\text{NF}}$ implements the next-fit policy using the "roving pointer" technique proposed in [19]: a global variable points to the chunk in the free-list involved in the last allocation or deallocation; `malloc` searches for a fitting free chunk starting from this pointer. Thus, the next-fit policy is a first-fit from the roving pointer. DK$_{\text{NF}}$ is challenging because the roving pointer introduces a case splitting that increases the size (number of disjuncts) in abstract values. The KR allocator [17] keeps a circular singly linked list, circularly sorted by the chunk start addresses; the start of the free-list points to the last deallocated block. The circular shape of the list requires to keep track of the free chunk with the biggest start address and this increases the size of the abstract values.

The analysis of each example with client program in Fig. 2 takes less than 3 seconds on a 2.53 GHz Intel Core 2 Duo laptop with 4GB of RAM. The universally quantified invariants inferred for DMA policies are given in the last column. Columns $|a^\sharp|$ and $|W_H|/|W_F|$ provide the maximum number of disjuncts generated for an abstract value resp. the maximum number of predicate atoms in each abstraction level.

---
[3] https://www.irif.univ-paris-diderot.fr/~sighirea/celia/plus.html

## 7   Related Works and Conclusion

Our analysis infers complex invariants of free-list DMA implementations due to the combination of two ingredients: the hierarchical representation of the shape of the memory region managed by the DMA and an abstract domain for the numerical constraints based on universally quantified formulas. The abstract domain has a clear logical definition, which facilitates the use of the inferred invariants by other verification methods.

The proposed abstract domain extends previous works [3, 5, 10, 11, 21]. We consider the SL fragment proposed in [5] to analyse programs using pointer arithmetic. We enrich this fragment in both spatial and pure formulas to infer a richer class of invariants. E.g., we add a heap-list level to track properties like chunk overlapping and universal constraints to infer first-fit policy invariants.

The split of shape abstraction on levels is inspired by works on overlaid data structures [11, 21]. We consider here a specific overlapping schema based on set inclusion which is adequate for DMA. We propose new abstract transformers which do not require user annotations like in [21]. Another hierarchical analysis of shape and numeric properties has been proposed in [25]. They consider the analysis of linked data structures coded in arrays and track the shape of these data structures and not the shape of the free set. Their approach is not based on logic and the invariants inferred on the content of list segments are simpler.

Our abstract domain includes a simpler version of the data word domain proposed in [3, 10], since the universal constraints quantify only one position in the list. Several abstract domains have been defined to infer invariants over arrays, e.g., [13] for array sizes, [14, 15] for array content. These works infer invariants of different kind on array partitions and they can not be applied directly to sequences of addresses. Recently, [22] defined an abstract domain for the analysis of array properties and applies it to the Minix 1.1 DMA, which uses chunks of fixed size. A modular combination of shape and numerical domains has been proposed in [6]. We extend their proposal to combine shape domains with domains on sequences of integers. Precise analyses exist for low level code in C [23] or for binary code [2]. They efficiently track properties about pointer alignment and memory region separations, but can not infer shape properties.